1

# Heavy neutrinos, Z' and Higgs bosons at the LHC: new particles from an old symmetry

S. Khalil[1,,32] and S. Moretti[3,4]

[1]*Center for Theoretical Physics, Zewail city of science and technology, Sheikh Zaid, 12588 Giza, Egypt.*

[2]*Department of Mathematics, Faculty of Science, Ain Shams University, Abbassia, 11566, Cairo, Egypt.*

[3]*School of Physics and Astronomy, University of Southampton, Highfield, Southampton SO17 1BJ, UK.*

[4]*Particle Physics Department, Rutherford Appleton Laboratory, Chilton, Didcot, Oxon OX11 0QX, UK.*

**A new era in particle physics is being spurred on by new data from the Large Hadron Collider. Non-vanishing neutrino masses represent firm observational evidence of new physics beyond the Standard Model. An extension of the latter, based on a $SU(3)_C \times SU(2)_L \times U(1)_Y \times U(1)_{B-L}$ symmetry, incorporating an established Baryon minus Lepton number invariance, is proposed as a viable and testable solution to the neutrino mass problem. We argue that LHC data will probe all the new content of this model: heavy neutrinos, an extra gauge boson emerging from spontaneous breaking of the additional gauge group at the TeV scale, onset by a new heavier Higgs boson, also visible at the CERN proton-proton collider. An even more exciting version of this model is the one exploiting Supersymmetry: firstly, it incorporates all its well known benefits; secondly, it alleviates the flaws of its more minimal realisations. Finally, this model provides a credible cold Dark Matter candidate, the lightest sneutrino, detectable in both underground and collider experiments.**

The search for new physics beyond the Standard Model (SM) at TeV energy scales is the major goal of the Large Hadron Collider (LHC) experiments at CERN. The LHC is a proton-proton collider which will unravel the next layer of fundamental physics [1]. It is currently running with a centre-of-mass energy of 7 TeV and a luminosity of $2 \times 10^{32}$ cm$^{-2}$ s$^{-1}$ and is expected to reach the final values of 14 TeV and $10^{34}$ cm$^{-2}$ s$^{-1}$, respectively. Therefore, possible new physics beyond the SM which is predicted to occur at the TeV energy scale can certainly be explored herein. The necessity of the latter is clear from experiment.

The SM of elementary particle physics, while in stunning agreement with all existing experimental results till a few years ago, we now know that is failing in one crucial respect. The evidence for non-vanishing neutrino masses, based on the observation of neutrino oscillations, indicates in fact that the SM requires an extension, as its left-handed neutrinos (denoted by $\nu_L$) are strictly massless due to the absence of right-handed neutrinos (denoted by $\nu_R$) and an exact global Baryon minus Lepton (B-L) number conservation.

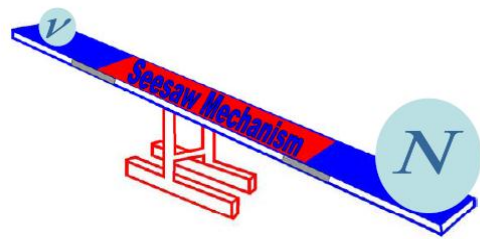

**Figure 1: Light neutrino mass generation via the seesaw mechanism.**

The most attractive mechanism that can naturally account for small yet sizable neutrino masses is known as the seesaw mechanism [2] (Fig. 1). In this case, three heavy singlet (right-handed) neutrinos are invoked, with super-heavy masses, of order $10^{13}$ GeV. Although this dynamics explains in a rather elegant way why neutrinos are much lighter than the other elementary fermions, it has no direct low energy signature.

In exploring the possibility of new physics beyond the SM, two key aspects should be borne in mind [3]. On the one hand, the very precise experimental confirmations of the SM that have kept accumulating throughout the last four decades make it mandatory for any theory of new physics to exactly reproduce the SM up to the Electro-Weak (EW) energy scale, of O(100 GeV). On the other hand, the tremendous success of gauge symmetry in describing nature, which the SM relies upon, implies a general belief that any new physics should be based on an enlarged gauge group. The simplest possible such an extension of the SM is the one based on the gauge group $SU(3)_C$ x $SU(2)_L$ x $U(1)_Y$ x $U(1)_{B-L}$ [4]. As mentioned, the SM is characterised by possessing a global $U(1)_{B-L}$ symmetry. If this symmetry is locally gauged, then the existence of three SM singlet fermions (the aforementioned right-handed neutrinos, $\nu_R$) is a quite natural assumption to make in order to cancel the associated anomaly, which is a necessary condition for the consistency of the model [5,6].

In addition, the model also contains an extra gauge boson (hereafter denoted by $Z'_{B-L}$), corresponding to the broken B-L gauge symmetry, and an extra SM singlet scalar, responsible for it, which is heavier than its SM counterpart. In general, the scale of B-L symmetry breaking is unknown, ranging from O(TeV) to much higher scales. However, it has been proven in [7] that, in a SUSY framework [8], the energy scale of B-L breaking is naturally correlated with that of soft SUSY breaking, which is indeed at TeV energies. Therefore, all these new particles will lead to novel signatures at experiments probing such an energy regime.

After $U(1)_{B-L}$ symmetry breaking has occurred, right-handed neutrinos acquire a mass $M_R = \lambda_R v'$, where $\lambda_R$ is a Yukawa coupling of order one and $v'$ is the Vacuum Expectation Value (VEV) onsetting B-L symmetry breaking, which, as mentioned, can be of O(TeV). Once standard EW Symmetry Breaking (EWSB) has also occurred, a Dirac neutrino mass $m_D = \lambda_v v$ is finally generated. (Here v is the VEV responsible for EWSB in the SM and $\lambda_v$ is a Yukawa coupling.) Therefore, the mass of the physical light neutrino (denoted by $\nu_l$) is given by $m^2_D/M_R$, which can account for the measured experimental results on neutrino oscillation if $\lambda_v \sim 10^{-6}$ [5]. While obviously small, the latter value is clearly not unnatural, as, e.g., the Yukawa coupling of the electron is $\sim 10^{-5}$. Despite their smallness, such couplings induce new interaction terms between, in particular, the physical heavy neutrinos (denoted by $\nu_h$), the associated leptons and the weak gauge bosons of the SM (W and Z). In addition, in Ref. [9], it was shown that a TeV scale B-L model can also accommodate another scenario in generating light neutrino masses, known as the inverse seesaw mechanism. This scenario was based on pioneering work in Ref. [10]. In this case, Yukawa couplings are no longer suppressed and can be of order one. Either way, in both these realisations of the seesaw mechanism, the heavy neutrinos associated to the B-L model are quite accessible and have interesting phenomenological implications.

In particular, at the LHC, the lightest heavy neutrinos can be (pair) produced via Z'$_{B-L}$ exchange (Fig. 2). (Notice that, in the case of the SM trivially extended with right-handed neutrinos only and no gauge modifications, their production is mainly in single mode through the exchange of a W boson, process which is very suppressed by the small mixing between light and heavy neutrinos.) Furthermore, the main decay channel of $\nu_h$ pairs is through two W bosons. Possible very clean signals, which would enable reconstruction of both the $\nu_h$ and Z'$_{B-L}$ masses, are those involving: (i) two pairs of charged leptons and missing transverse energy (due to two light neutrinos escaping detection) [11]; (ii) three charged leptons, two jets and missing transverse energy (due to a single undetected light neutrino) [6].

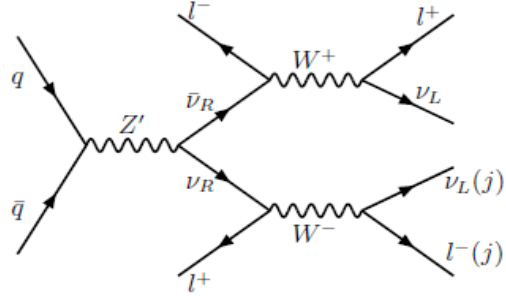

**Figure 2:** Two di-leptons (Tri-lepton and di-jet) plus missing transverse energy signature of $\nu_R$ pairs produced at the LHC.

A further exciting possibility for the right-handed neutrinos is to be long-lived particles, with such final states embedding then distinctive displaced vertices, in which a highly energetic and isolated pair of leptons point to a different vertex than the primary one. In essence, the aforementioned signatures can provide a powerful insight in the whole leptonic sector of the B-L model (and into its interplay with the gauge one as well) as the allow for not only the measurement of the heavy neutrino masses but also of their lifetimes (through the displacement length of their decay vertices). Further, from the simultaneous measurement of these two quantitites, even the light neutrino masses can be inferred.

The Higgs sector in the B-L model consists of the SM complex Higgs doublet and a further complex Higgs singlet. Out of the associated six (scalar) degrees of freedom, only two survive in the form of physical objects after the B-L and EW symmetries are broken. The other four degrees of freedom are `eaten' by the Z'$_{B-L}$, Z and two W bosons. The superposition between the two Higgs scalar fields is controlled by their mixing parameter in the scalar potential. In terms of mass eigenstates, we find one light Higgs, H, and a heavy one, H', with a mixing angle $\theta$. Due to this mixing between the two Higgs bosons, the usual couplings among the SM-like Higgs, H, and the SM fermions and gauge bosons are modified (essentially multiplied by $\cos\theta$). In addition, new couplings among the extra Higgs state, H', and the SM particles are obtained (proportional to $\sin\theta$) [13]-[15].

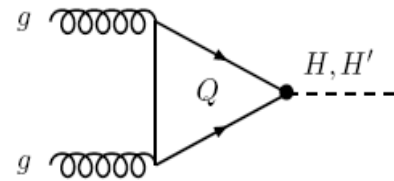

**Figure 4:** Diagram for H and H' production at the LHC.

At the LHC, the dominant channel for Higgs boson production is due to gluon-gluon fusion, which is mediated by triangular loops of heavy quarks (see Fig. 4). Thus, the cross section of this process is proportional to the Higgs boson couplings to the heavy quark mass. In case of the B-L extension of the SM, one notices that the production cross section for the light Higgs state is reduced respect to the SM one by a factor $\sim\cos^2\theta$. Conversely, heavy Higgs production is suppressed by two effects: a small $\sim\sin^2\theta$ coupling and a large $m_{H'}$ (compared to $m_H$). Therefore, heavy Higgs production rates are typically less than those of the light Higgs yet still sizable [15]. In this class of models, in addition to SM-like decay channels, either or both Higgs bosons can decay in genuine B-L final states, like $v_h$ and/or $Z'_{B-L}$ pairs, with sizable rates. This opens up then the intriguing possibility of all the new states predicted by the B-L model being *simultaneously* detected at the LHC [15] (see also [16]-[18]).

Furthermore, let us mention that, in the pursuit of a grand unified theory based on a B-L extension of the SM manifesting itself at the TeV scale, an unavoidable `rite of passage' is the `Supersymmetrisation' of the model. While naturally acquiring all the remedies offered by a Supersymmetric (SUSY) scenario against SM flaws (like absence of gauge coupling unification and the hierarchy problem, which would persist in a non-SUSY B-L extension), a SUSY version of the latter would in particular alleviate even known drawbacks of the minimal SUSY model (i.e., the one without heavy neutrinos, $Z'_{B-L}$ boson and singlet Higgs state plus SUSY counterparts). A striking example is the case of a SUSY B-L model with inverse seesaw, whereby, as shown in [19], the one-loop radiative corrections to the lightest SM-like Higgs boson mass, due to the right-handed neutrinos and sneutrinos can give an absolute upper limit on it at around 170 GeV. This enhancement greatly reconciles theory and experiment, by alleviating the so-called `little hierarchy problem' of the aforementioned minimal SUSY realisation, whereby the current experimental limit on the SM-like Higgs mass of 115 GeV is very near its absolute upper limit predicted theoretically, of 130 GeV or so. Conversely, a SM-like Higgs boson with mass below 170 GeV is still well within the reach of the LHC, so that the SUSY B-L realisation discussed here is just as testable as the minimal version. Recent experimental hints of Higgs boson signals at the LHC are therefore fully compatible with both non-SUSY and SUSY versions of the B-L scenarios we discussed.

In short, a symmetry structure deeply rooted in the SM could well be the key to extend the latter into a credible new physics scenario, embedding naturally the neutrino mass patterns measured by experiment and at the same time offering a wealth of new physics signals, all promptly accessible at the LHC, as the dynamics generating the new states occurring in the model emerges at the TeV scale (and particularly so in its SUSY versions), hence well within the reach of the CERN collider.

**Acknowledgements**

The work of S.M. is funded in part by the NExT Institute and in part by the STFC (Swindon, UK). The work of S.K. is partially supported by the Leverhulme Trust under the grant VP2-2011-012. A special thank goes to all our collaborators.